\documentclass[sigconf,screen]{acmart}
\AtBeginDocument{%
  }

\setcopyright{acmlicensed}
\copyrightyear{2018}
\acmYear{2018}
\acmDOI{XXXXXXX.XXXXXXX}
\acmConference[Conference acronym 'XX]{Make sure to enter the correct
  conference title from your rights confirmation email}{June 03--05,
  2018}{Woodstock, NY}
\acmISBN{978-1-4503-XXXX-X/2018/06}
\graphicspath{{figs/}{figures/}{pictures/}{images/}{./}} 
\usepackage{cleveref}
\usepackage{url}
\usepackage{tabu}                      
\usepackage{booktabs}                  
\usepackage{lipsum}                    
\usepackage{mwe}                       
\usepackage{newtxtext}
\usepackage{newtxmath}
\usepackage{color}
\usepackage{xcolor}
\usepackage{array}
\usepackage{enumitem}
\usepackage{makecell}
\usepackage{listings}

\usepackage{xspace,xpunctuate}

\newcommand{\etal}{\xspace\textit{et al.}\xspace}
\newcommand{\eg}{\textit{e.g.},\xspace}

\begin{document}
\graphicspath{{figs/}{figures/}{pictures/}{images/}{./}} 



\title{SenseWalk: Agent-Based Semantic Trajectory Simulation Powered by Large Language Models in Zoned Environments}

\author{Ziyue Lin}
\email{ziyuelin917@gmail.com}
\orcid{0009-0002-5485-7379}
\affiliation{%
  \institution{School of Data Science, Fudan University}
   \streetaddress{1 Th{\o}rv{\"a}ld Circle}
  \city{Shanghai}
  \country{China}
}

\author{Xinhang Xie}
\affiliation{%
  \institution{School of Data Science, Fudan University}
   \streetaddress{1 Th{\o}rv{\"a}ld Circle}
  \city{Shanghai}
  \country{China}
}
\author{Kangyi Wang}
\affiliation{%
  \institution{School of Data Science, Fudan University}
   \streetaddress{1 Th{\o}rv{\"a}ld Circle}
  \city{Shanghai}
  \country{China}
}
\author{Siming Chen}
\authornote{Siming Chen is the corresponding author.}
\email{simingchen@fudan.edu.cn}
\orcid{0000-0002-2690-3588}
\affiliation{%
  \institution{School of Data Science, Fudan University}
   \streetaddress{1 Th{\o}rv{\"a}ld Circle}
  \city{Shanghai}
  \country{China}
}


\renewcommand{\shortauthors}{Trovato et al.}

\begin{abstract}
  Semantic trajectory analysis has recently emerged as an approach for modeling human movement by capturing implicit patterns and behaviors through semantic information (\eg visitors' profiles and goals) beyond raw spatial paths to better understand why people move in certain ways. 
  However, analyzing semantic trajectories in real-world scenarios remains challenging, as collecting high-quality data is costly and often lacks rich semantic information.
  Meanwhile, existing simulation tools require substantial technical expertise, which makes them difficult for practitioners to adopt.
  To address these limitations, the paper proposes ${SenseWalk}$, an interactive system that supports simulating semantic trajectories by LLM-powered agents.  
  We develop a simulation workflow that combines LLMs and the social force model to balance physical plausibility and semantic coherence.
  A user-friendly interface is designed to facilitate users in customizing the simulation configuration and analyzing simulation outputs.
  We also conduct a quantitative experiment to evaluate the effectiveness of our simulation workflow, and a user study (n=12) to assess the usefulness and efficiency of our system.
\end{abstract}



\received{20 February 2007}
\received[revised]{12 March 2009}
\received[accepted]{5 June 2009}

\maketitle
\section{Introduction}
Imagine a museum professional seeking to understand how different types of visitors navigate and engage with exhibitions.
For instance, a first-time tourist may follow on-site signage and focus primarily on iconic artworks, whereas a local art student who frequently visits might adopt a self-directed route, selectively revisiting favored sections.
Analyzing these distinct patterns holds substantial promise for personalized route recommendation~\cite{recommend_2021,target_2019}, adaptive exhibition design~\cite{design_2024}, and nuanced engagement evaluation~\cite{experiment_2024}, which are increasingly important for cultural institutions and public venues~\cite{semantic_museum_2021,generator2024he,recommend_2021,hybrid2025zuo}.

Focusing solely on raw spatial paths or aggregate footfall statistics is insufficient to explain visitors’ movement. Although the two visitors may traverse overlapping areas, their engagement unfolds differently. 
To capture these evolving patterns, prior work introduced the concept of semantic trajectories~\cite{Semantic_model_2013,semantic_prediction_2011,mining2010ying}, which augment spatial and temporal coordinates with semantic annotations describing the trajectory in its entirety.
Semantic annotations are not limited but can be broadly categorized into two main types: behaviors and goals~\cite{semantic_museum_2021,warehouse2019michael}.
Behavior annotations reflect the actuality of movement~(\eg wandering around an exhibition), whereas goal annotations indicate the potentiality of movement~(\eg aiming to visit the Mona Lisa), even if the goal is not ultimately achieved.

However, collecting semantic trajectory data is operationally demanding. First, high-quality spatiotemporal tracking often requires costly and complex sensing infrastructure~(\eg sensors, wearables, or multi-camera systems) and still suffers from noise and coverage gaps~\cite{position2024hailu,track2025auro}. 
Second, semantics can not be directly observed. 
Enriching raw trajectories into semantic representations typically relies on post-hoc annotation techniques, which require additional contextual sources and domain expertise~\cite{enrichment2023lettich,semitri2011yan}. 
Third, visitor profiles that associate with trajectories are difficult to obtain. Attributes such as age, interests, or time budget are rarely available at scale, and collecting them often relies on intrusive surveys, which raises privacy and consent concerns.
Finally, many real-world settings face a cold-start scenario with little or no historical visitor trajectory data. For instance, when a new exhibition is launched, prior visitor trajectories are unavailable, yet route guidance must still be made.

Motivated by the limitations, we aim to develop a simulation tool for trajectory-aware practitioners who make decisions based on visitor movements to generate semantic trajectories tailored to their needs. 
Our work focuses on zoned environments~\cite{pena2012problem,syntax2021yamu}~(\eg museum, exhibition, and shopping mall), which are spatially bounded areas intentionally divided into semantically distinct zones.
There are three major challenges to overcome.
First, simulating semantic trajectories requires balancing physical plausibility and semantic coherence.
Simulations must respect spatial constraints, such as accessibility and movement costs, while also reflecting visitors' characteristics and the semantic logic of the visit.
Second, enabling effective and flexible authoring remains challenging, as many target users lack technical or modeling expertise.
During practitioners’ analytic workflow, they need to repeatedly adjust assumptions and explore alternative scenarios to converge on desired outcomes.
Existing commercial tools like Anylogic\footnote{\url{https://www.anylogic.com/}} and PTV Viswalk\footnote{\url{https://www.ptvgroup.com/en/products/pedestrian-simulation-software-ptv-viswalk}} primarily entail a steep learning curve, requiring familiarity with parameter calibration and engineering workflows.
The complexity limits the practical usability of such tools in real-world applications.
Third, interpreting generated trajectories is nontrivial. In practice, practitioners need to understand not only where visitors move, but also why specific decisions occur, such as visiting or skipping particular zones, to build trust and translate simulation outputs into actionable insights.

To address the challenges, we present ${SenseWalk}$, an interactive system for simulating semantic trajectories with LLM-powered agents. 
To build the simulation workflow of agents, we integrate the social force model~\cite{force2010mou} to enforce physical plausibility with the LLMs' language understanding to translate high-level semantic constraints.
The system enables practitioners to customize environment maps and agent profiles for alternative scenario exploration and to analyze the trajectories through intuitive interaction design.
We visualize the simulated semantic trajectories, including geographical positions and semantic annotations.
We also support practitioners chatting with agents to give them instructions and to access their internal reasoning, which enables users to understand why agents make certain decisions.
We conduct an experiment to assess the effectiveness of our simulation workflow and a user study to verify the usefulness and efficiency of our system.
In summary, our contributions include:
\begin{itemize}[leftmargin=*,topsep=0pt]
    \item A simulation workflow for semantic trajectories that combines LLMs and social force models to balance physical plausibility and semantic coherence. 
    \item A user-friendly and intuitive interactive system that lowers the barrier to support authoring alternative scenarios and analyzing trajectories.
    \item A quantitative experiment to evaluate the effectiveness of the simulation workflow and a user study to verify the usefulness and efficiency of our system.
\end{itemize}
\section{Related Work}
We discuss related studies, including semantic trajectory modeling and analysis, LLM-based agent simulation, and agent-based modeling and simulation for decision-making.
\subsection{Semantic Trajectory Modeling and Analysis}
Motivated by the need to move beyond raw movement points toward interpretable representations, the semantic trajectory has been extensively studied in the mobility communities~\cite{giannotti_unveiling_2011,mining2007gian,engine2008gko}.
The trajectory is identified by two specific spatio-temporal positions of the movement track~\cite{SPACCAPIETRA2008126}.
Semantic trajectory enriches these coordinates with contextual annotations to enhance the analysis of movement data and facilitate the discovery of semantically implicit behavior patterns~\cite{Semantic_model_2013, survey2015albanna}.
A promising enrichment approach is to leverage ontologies and linked data to semantically connect trajectory patterns with broader behavioral and contextual knowledge~\cite{recommend_2021}.
Semantic trajectory analysis has broad practical applications. 
For example, Kontarinis\etal~\cite{semantic_museum_2021} incorporate semantic information to support personalized route recommendations, enhancing visitor experiences in museums.
RetailOpt~\cite{Retail2024yone} tracks customer motion data and purchase records in indoor retail environments for customer behavior analysis and in-store navigation.

Due to data shortage and privacy restrictions, researchers have focused on trajectory simulation methods, which can be categorized into rule-based~\cite{LAMMEL2015950} and data-driven~\cite{rasouli2022pedformerpedestrianbehaviorprediction,hybrid2025zuo}.
Rule-based models encode the walking trajectories through predetermined mechanisms, such as mathematical equations~\cite{integrateXi2011} or cellular automata~\cite{seitz2016superposition}.
Data-driven simulation models use deep neural networks to generate trajectories by learning from the training data~\cite{Zhang_2020_CVPR,graph2022fang,cao2025holisticsemanticrepresentationnavigational}.
However, existing methods struggle to capture the deeper semantic logic underlying movement, resulting in annotations that label where movement occurs but offer limited understanding of why it occurs.
We utilize LLMs to serve as a semantic annotation engine that bridges low-level trajectories and high-level interpretations.
The context-awareness enables it to align annotations with contextual inputs, such as zone descriptions, visitor profiles, temporal constraints, and environmental semantics.
LLMs can also provide flexibility beyond predefined rule-based annotation types and dynamically adapt the granularity to suit diverse practitioner needs.

\subsection{LLM-Based Agent Simulation}
LLM-based agents have recently emerged as a practical paradigm for modeling human-like behavior~\cite{wang_survey_2024}. 
Generative Agents~\cite{generative2023park} demonstrated an LLM-centered architecture that combines memory, planning, and reflection to produce believable daily routines and social interactions.
Sensible Agent~\cite{sensible2025lee} and InterQuest~\cite{inter2025mei} studied the agent's autonomy to model user intents proactively.
SketchGPT~\cite{sketch2025huang} extends agents’ perceptual capabilities by enabling them to interpret not only textual input but also sketches and speech, supporting richer multimodal interaction. 
Given the exceptional reasoning and planning capabilities of LLMs, multi-agent collaboration systems, in which multiple autonomous agents coordinate their actions to achieve shared goals, have been developed in various domains~\cite{li_survey_2024}.
PosterMate~\cite{poster2025shin} supported a collaboration mechanism among agents with diverse capabilities to assist poster design.
MetaGPT~\cite{hong2024metagpt} assigned diverse roles and communication rules to various agents for automatic programming tasks.
StorySage~\cite{story2025tala} utilized a multi-agent framework for autobiography writing of diverse users.

Human mobility modeling has also been a promising direction of LLM-based agents, which shift from purely text generation toward geographic synthesis~\cite{mobile2024gong}. 
GATSim~\cite{liu2026gatsimurbanmobilitysimulation} and OpenCity~\cite{yan2024opencityscalableplatformsimulate} utilized language understanding capabilities of LLMs to simulate urban mobility based on diverse agent profiles, lifestyles, and preferences.
Wang\etal~\cite{urban2024wang} proposed a LLM-agent framework that accounts for individual activity motivations for personal mobility
generation.
PuppetLine~\cite{puppet2025wang} decomposed text input into trajectories and emotions by leveraging a library of action primitives primitives.
Although LLMs excel at generating contextually plausible movement decisions, they often exhibit limited sensitivity to spatial constraints, such as connectivity, accessibility, and movement costs.
Ju\etal~\cite{trajllm2025ju} combine LLMs with physical models for location mapping to generate coherent human mobility; however, the approach does not explicitly model fine-grained pedestrian dynamics under continuous spatial constraints.
To bridge the gap, our paper introduces a hybrid algorithm to integrate LLMs and the social force model that enforces physically plausible motion and collision-free navigation.

\subsection{ABMS for Decision Making}
Agent-based modeling and simulation~(ABMS)~\cite{abms2009macal} served as a core methodology for human behavior research and provides decision support for scenarios that are difficult to test in the real world~\cite{ross2022simulation}.
ABMS models complex systems composed of autonomous agents whose local rules produce emergent collective behavior~\cite{abms2005macal}, which has been applied across diverse domains, including finance~\cite{finance2025axtell}, pandemic~\cite{nitzsche_agent-based_2024}, and sociology~\cite{sociology2015bian}.
It has further been adopted to simulate trajectories in structured environments.
Rahbar\etal~\cite{arch2022mor} applied ABMS to generate human trajectories to facilitate architectural layout designs.
Gabilondo\etal~\cite{retail2024san} synthesized mobility data to model non-residential demand in Tokyo’s commercial streets.

Human-AI interactions engaged with simulations provide a human-in-the-loop way to shape and interpret simulated behavior, allowing users to iteratively explore scenarios and test assumptions while observing how outcomes unfold~\cite{lin2025carbonsiliconcoexistcompete}.
Commercial platforms such as NetLogo\footnote{\url{https://www.netlogo.org/}} and Swarm\footnote{\url{http://www.santafe.edu/projects/swarm/}} enable researchers to specify behavioral rules and dynamically adjust parameters to observe emergent collective phenomena.
However, these systems typically require programming expertise and explicit rule specification, resulting in steep learning curves and limiting their accessibility and applicability across diverse real-world scenarios.
LLMs have broadened ABMS by enabling more natural human–AI interaction and lowering barriers for non-experts to communicate with simulations via natural language~\cite{virt2025almu,society2025zhang}.
DialogLab~\cite{dialog2025hu} supported authoring and simulating interactively multi-party dialogues.
Social Simulacra~\cite{social2022park} and SimSpark~\cite{simspark2025lin} provide prototypes for generating social media behaviors according to users' designs.
Kim\etal~\cite{peer2025kim} simulate a deliberative agent that induces socio-cognitive conflict and communicates with human users to foster democratic skills.
Inspired by existing work, our interactive system aims to provide a user-friendly platform for trajectory-aware practitioners. 
Users are allowed to simulate semantic trajectories by intuitive interactions and natural language, reducing the learning curve. 
\section{Preliminary Study}
To figure out the design requirements of our interactive system, we conducted semi-structured interviews with practitioners to obtain valuable insights.
\subsection{Participants and Procedure}
We recruited 7 participants~(4 males, 3 females) who are our system's target users, with ages ranging from 26 to 41 years old~($Age_{mean} = 32.86$, $Age_{std} = 5.58$).
Participants comprised a diverse group of trajectory-aware practitioners working in zoned environments, such as museums, expos, and similar settings.
They all had substantial professional experience and were deeply familiar with trajectory analysis and operational decision-making.
Their analysis goals are practice-driven, including route recommendation, engagement evaluation, layout optimization, visitor guidance, and promotion placement.
Meanwhile, most participants reported low to moderate familiarity with simulation tools, indicating that even highly experienced professionals may still face barriers when adopting existing simulation-based approaches.
Detailed demographic information is shown in~\Cref{tab:predemogra}.

We obtained informed consent from participants to record the interviews.
Each interview lasted approximately 45-60 minutes. 
First, we interviewed participants about their current practices in analyzing trajectory data, including their typical analytical workflows, the perceived importance of trajectory analysis for decision making, and the key challenges they encounter.
Next, we asked about their experiences with and opinions on existing simulation tools.
After introducing our preliminary idea of an LLM-powered agent simulation system, we invited participants to share their expectations and suggestions for such a system.
Insights from these stages informed the design requirements.

\subsection{Findings}
We summarize four key findings~(\textbf{F1-F4}) from the interview.

\textit{\textbf{F1: Challenges of obtaining real data.}} 
Participants agreed that acquiring real-world trajectory data is costly and often impractical.
Collecting high-quality traces and visitors' profiles typically requires substantial sensing infrastructure and demanding data preprocessing, while semantic information, such as visitors’ goals, is even harder to capture directly.
\textbf{P4} noted \textit{``Collecting this kind of data at scale requires too much coordination and time. We sometimes collect visitor information through questionnaires, but the data quality is often low, and the sample size is limited.''}
\textbf{P3} explained \textit{``We can know where visitors go, but it is much harder to know why they make those choices.''}
In addition, privacy and ethical concerns further restrict access to fine-grained data, especially in public venues.

\textit{\textbf{F2: Complexities of Zoned Environments.}}
Zoned environments are often characterized by spatial concentration, with multiple functionally distinct zones densely arranged within a confined space, such as expos and museums.
\textbf{P1} and \textbf{P6} agreed that many different zones are packed into a relatively limited space.
These zones differ in a variety of attributes, including spatial properties (\eg position, size, and accessibility) and semantic properties (\eg theme and function). 
Such heterogeneity further shapes visitors’ decision-making, as movement is influenced not only by spatial proximity but also by the distinct characteristics of each zone.
Zoned environments also exhibit temporal dynamics.
\textbf{P4} noted \textit{``When a product-introduction talk or lucky draw starts, people suddenly gather there.''}

\textit{\textbf{F3: High Barrier to Simulation Tools.}}
Existing simulation tools present a high barrier to adoption in practice.
First, these systems with complex interfaces often require substantial technical expertise and sometimes programming skills. 
\textbf{P2} explained \textit{``It takes a great deal of time to learn how to use these tools properly, which can reduce our efficiency in everyday work.''}
Second, users must abstract real-world environments into model parameters, which is difficult, especially for users without modeling backgrounds.
\textbf{P6} said \textit{``It is hard to decide what values to use when we need to turn things like shop popularity or activity influence into numbers for the model''}

\textit{\textbf{F4: Contextualized Trajectory Interpretation.}}
Interpreting trajectories requires understanding the contextual factors that shape visitors’ decisions. 
Trajectories are often more informative when interpreted alongside contextual information, such as visitor profiles, goals, reasoning processes, and the semantic characteristics of different zones.
\textbf{P5} emphasized \textit{``A path alone does not tell us much unless we also know who the visitor is and what they are trying to do.''}
However, \textbf{P5} also pointed out the practical limitation, stating that \textit{``In real settings, it is very hard to know these details at scale, especially people’s goals or motivations.''}
A similar concern was echoed by \textbf{P2} \textit{``We may observe where visitors go, but it is much harder to capture why they make certain choices, or how they perceive different zones.''}

\subsection{Design Requirements}
After interviewing our target users and conducting literature reviews, we specified three design requirements:

\textbf{\textit{D1: Generate Plausible Semantic Trajectory.}}
The system should generate semantic trajectories that are both physically feasible and semantically coherent, thereby serving as a substitute for real-world data to facilitate more informed decision making~(\textbf{F1}).

\textbf{\textit{D2: Support Scenario Authoring.}}
The system should facilitate scenario authoring by enabling users to configure and modify simulation workflows without requiring extensive technical or modeling expertise~(\textbf{F3}).
It should further specify fine-grained configuration of scenario settings that closely reflect real-world contexts~(\textbf{F2}).

\textbf{\textit{D3: Facilitate Trajectory Analysis.}}
The system should support effective analysis of generated trajectories by enabling users to inspect movement patterns.
It should also help users understand the underlying reasons for these trajectories so that simulation results can inform planning and operational decisions~(\textbf{F4}).

\section{Simulation Workflow}
In this section, we propose the simulation workflow of semantic trajectories, involving simulation configuration and agent architecture.
The configuration defines the simulation scenario settings, whereas the architecture determines how agents act within that setup~(\textbf{D1}).
\subsection{Simulation Configuration}
\begin{figure}[t]
  \centering
  \includegraphics[width=\linewidth]{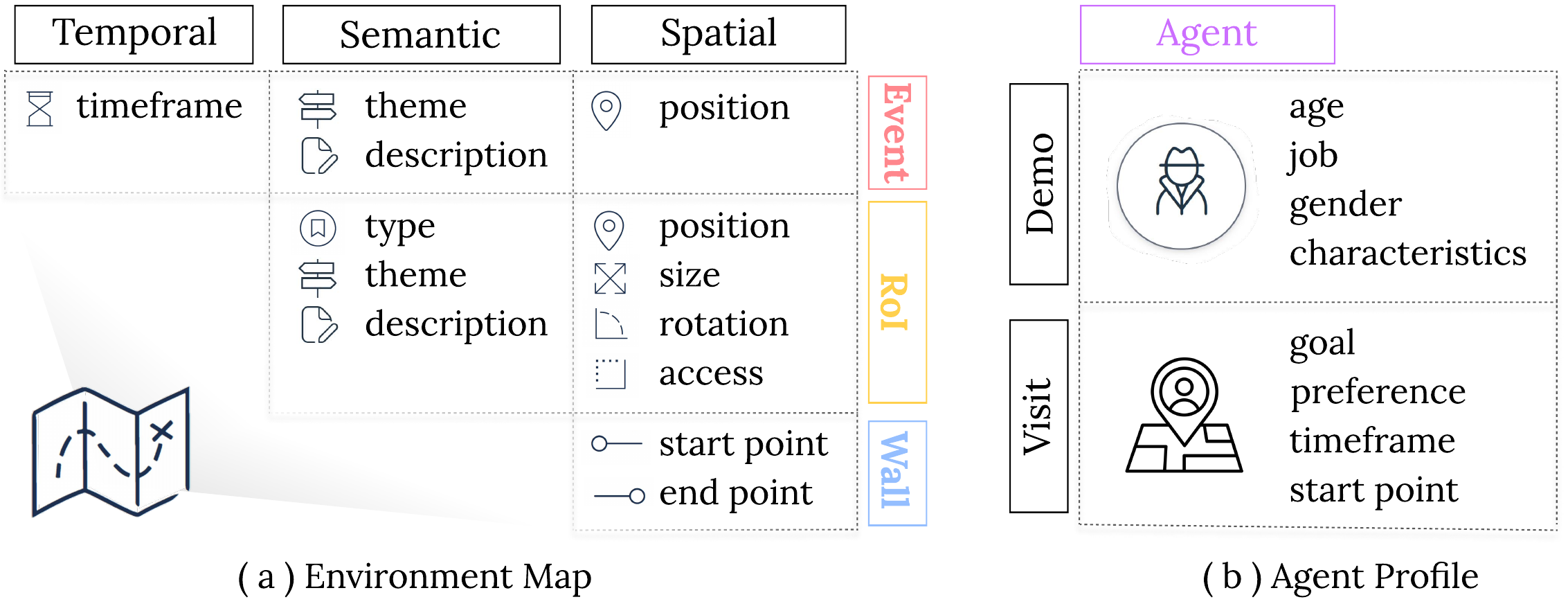}
  \caption{%
    \textbf{The simulation configuration} includes (a) Environment Map and (b) Agent Profile.
  }
  \label{fig:configure}
\end{figure}

Simulation configuration refers to the simulation workflow inputs tailored to the user's analysis requirements~(\textbf{D2}).
The configuration consists of two parts, environment maps and agent profiles~(\cref{fig:configure}).
\subsubsection{Environment Map}
Our preliminary study revealed that the environments in the target application scenarios are inherently complex~(\textbf{F2}).
We model the environment map by three components: walls, regions of interest~(RoIs), and events.
Walls specify the physical boundaries of the environment.
Each wall is represented as a line segment defined by two endpoints, which constrain walkable space and shape agents’ movement paths.
RoIs represent functional zones associated with specific semantic meanings.
For example, in an expo setting, RoIs may correspond to brand booths or a service center.
Each RoI is defined by spatial attributes, including its position, size, rotation, and access~(entrance sides), as well as semantic attributes such as theme, type, and detailed description.
Events introduce temporal dynamics into the environment.
Each event is associated with a spatial location and a temporal timeframe. 
It also includes semantic information: the theme and description.
\subsubsection{Agent Profile}
Users can customize agents' profiles that characterize each simulated visitor in the environment.
It includes basic demographic information, such as age, gender, job, and personal characteristics.
The profile also specifies behavioral attributes, including the agent’s visit goals and preferences, which influence decision-making and movement patterns.
In addition, each agent is associated with temporal attributes, namely the timeframe of their visit, as well as an initial spatial location that determines where the agent enters the environment.

\subsection{Agent Architecture}
\begin{figure}[t]
  \centering
  \includegraphics[width=\linewidth]{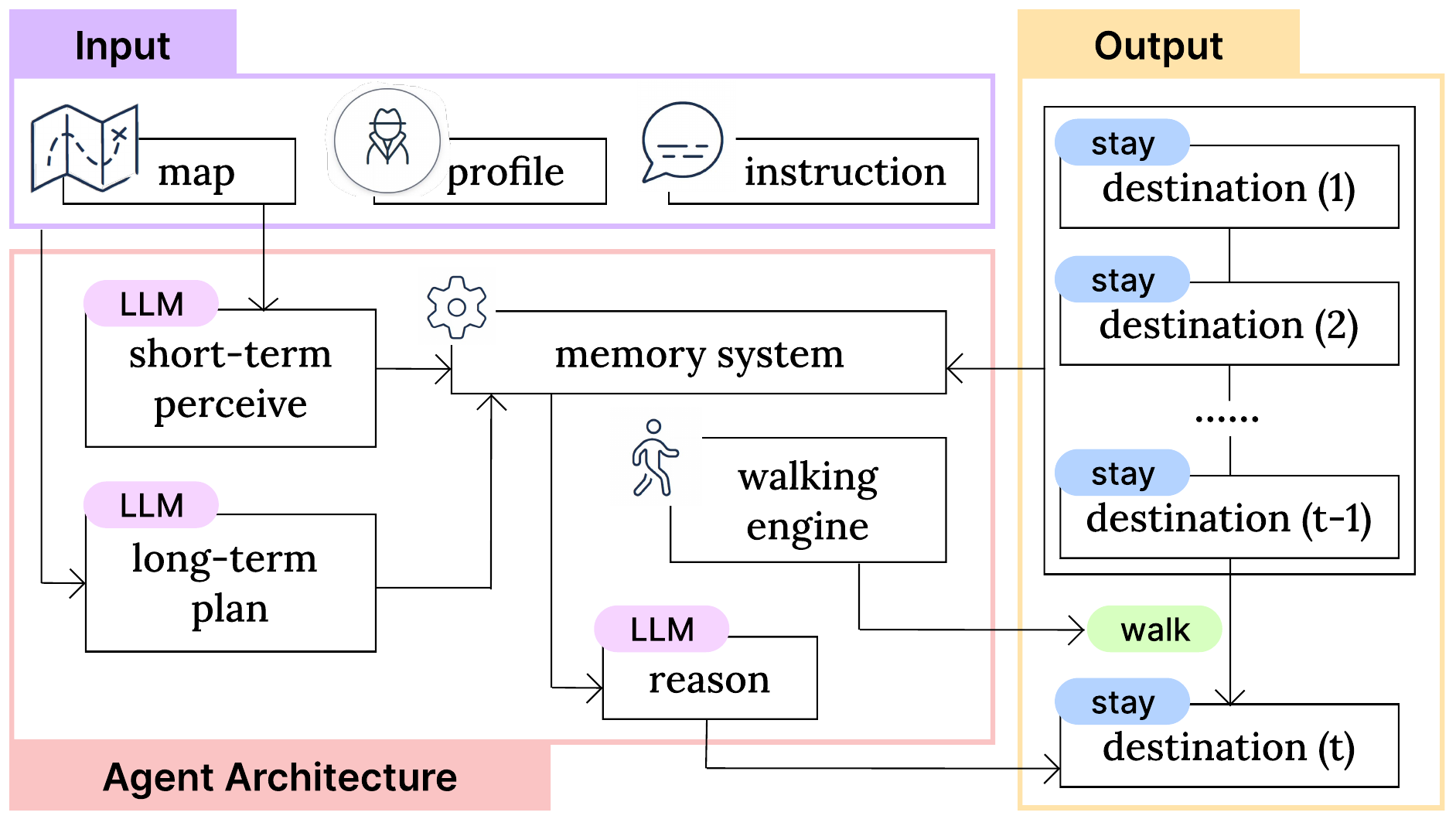}
  \caption{%
    \textbf{The simulation workflow.} Input consists of three parts: environment map, agent profiles, and user instructions. The agent architecture integrates multiple coordinated modules to generate geographic positions and semantic annotations as output semantic trajectories.
  }
  \label{fig:workflow}
\end{figure}

Agent architecture is powered by LLMs to generate semantic trajectories, including geographic positions and semantic annotations, based on simulation configuration~(\textbf{D1}).
It refers to the integrated decision-making mechanism that enables agents to interpret their simulated environment.
As illustrated in \Cref{fig:workflow}, it is composed of five coordinated modules.

First, the agents will formulate a visit \textbf{long-term plan} for simulating how people form intentions in advance in real-world settings.
For example, in an expo, a procurement buyer may typically make clear plans based on their purchasing needs, while a general visitor may instead plan a more exploratory route.
The module leverages simulation configuration: map and profile as input. 
In addition, users can provide explicit textual instructions to agents through the system.
We convert all inputs into a JSON format and utilize LLMs to generate a structured sequence of planned destinations, where each item specifies which zone~(RoI or event) the agent intends to visit, why it is selected, and when it is scheduled.
For each destination, an interest score is computed to represent how attractive the zone is, which is estimated by LLMs and scaled to the range of 0 to 1.

The trajectory is not solely determined by pre-defined plans, but is continuously influenced by nearby zones. We introduce the~\textbf{short-term perceive} module to enable agents to sense their surroundings within a fixed spatial range periodically.
For each nearby zone, an interest score is also computed.
If the score exceeds a predefined threshold, it will be recorded in the \textbf{memory system}.

We maintain a memory system for each agent to dynamically record its plans, perceptions, and visited zones.
Every time the memory system is updated, the agent will begin to \textbf{reason} about the next destination.
For each zone~(planned or percevied) in the memory system, we compute a score based on both distance and interest level.
The distance represents the spatial cost for an agent to reach a given zone and is normalized.
These two factors are then weighted and contrasted.
The highest-scoring unvisited zone is selected as the current destination.
When the agent reaches a destination, LLMs will be used to determine the duration of stay.

We develop a \textbf{walking engine} to simulate agents' certain movement.
While LLMs excel at generating semantically plausible decisions, they lack sensitivity to spatial constraints and often fail to account for factors such as collision avoidance.
Therefore, we adopt the social force model~\cite{force2010mou} to guarantee physical realism.
The social force model simulates pedestrian movement by a combination of virtual forces.
We use the following equation to compute agent $i$ motions:
\begin{equation}
\frac{d \vec{v}_i}{dt} = \vec{f}_i^{\,0} + \vec{f}_i^{\,\text{wall}},
\end{equation}
where $\frac{d \vec{v}_i}{dt}$ represents the acceleration of agents, the driving force $\vec{f}_i^{\,0}$ directs the agent toward its intended destination, and the repulsive force $\vec{f}_i^{\,\text{wall}}$ reflects the repulsive force from walls and inaccessible side of RoI is orthogonal to the wall.
When agents start to walk, they are guided by a driving force that pulls them toward their current destination.
In addition, when the agent approaches a wall or an inaccessible side of RoI within a certain threshold distance, it experiences a repulsive force from the wall, which pushes it away to prevent collisions.

\subsection{Workflow Overview}
After users configure the map and agent profiles, the workflow first generates a long-term plan consisting of a sequence of destinations for each agent. The agent then starts walking toward the first destination by walking engine. During movement, it periodically perceives its surroundings and reasons whether to adjust its destination as needed. Upon arrival, the agent reasons its stay time before proceeding to the next destination. The process repeats throughout the simulation. Meanwhile, the memory system is continuously updated to record the agent’s behaviors and reasoning process.
The final semantic trajectory output includes both the geographic positions and semantic annotations at each time step.
The social force model generates the geographic positions.
As mentioned, semantic annotations are classified into two types: goals~(current destination) and behaviors~(walking or staying)~\cite{semantic_museum_2021}.
Additionally, we provide all the prompt templates in~\Cref{sec:propmt}.

\section{Interface}
\begin{figure*}[t]
  \centering
  \includegraphics[width=\linewidth]{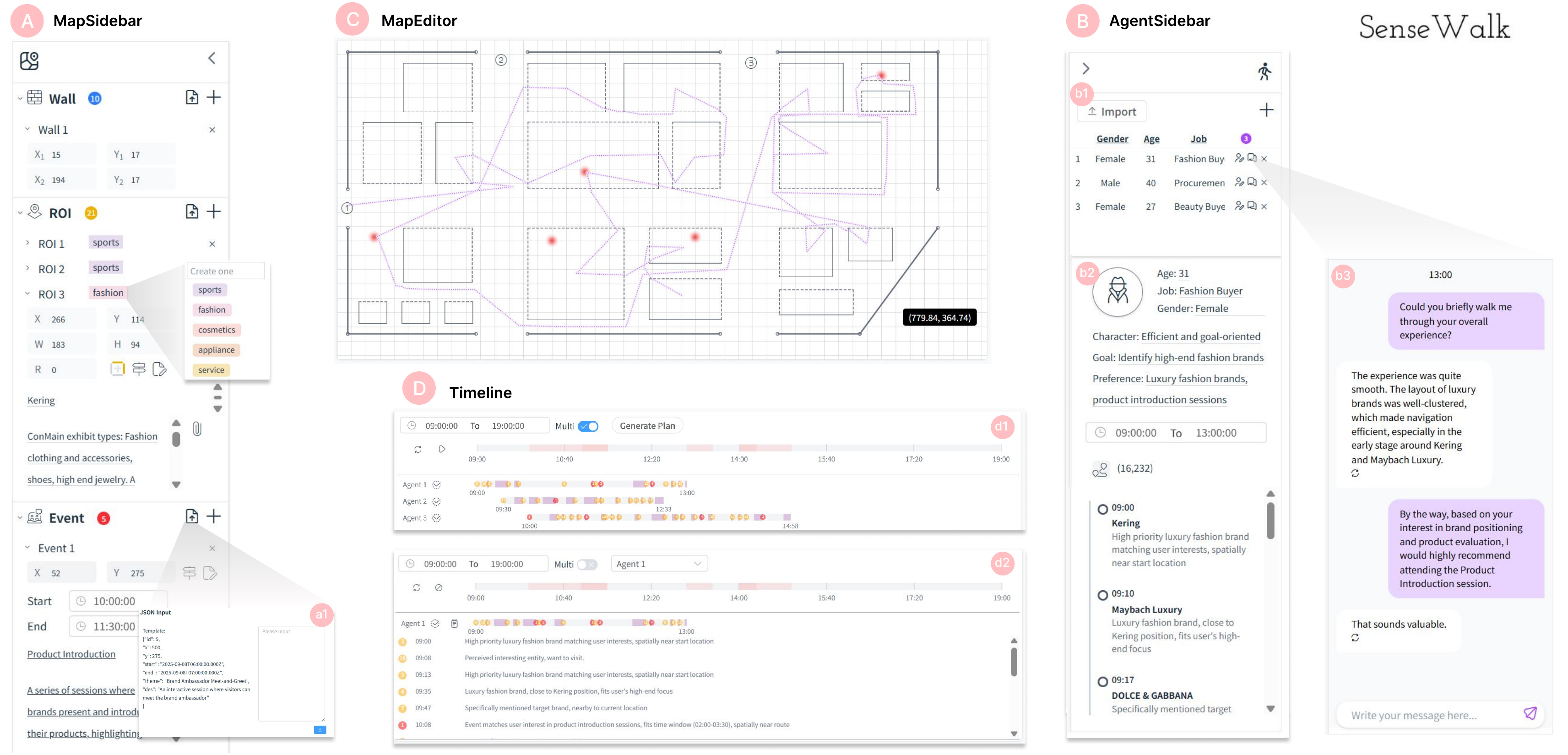}
  \caption{%
    \textbf{The interface of SenseWalk.} It consists of four views: (A) MapSidebar supports the environment map configuration; (B) AgentSidebar supports the agent profile configuration and allows users to chat with agents; (C) MapEditor displays the geographical positions of maps and trajectories; (D) Timeline monitors agents' behaviors and semantic annotations.
  }
  \label{fig:interface}
\end{figure*}

The user interface supports both authoring~(\textbf{D2}) and analyzing~(\textbf{D3}) stages, enabling practitioners to construct simulation scenarios and interpret the outcomes. 
It consists of four views: MapSidebar, MapEditor, Timeline, and AgentSidebar~(\Cref{fig:interface}).
The interface is implemented in TypeScript.
To better illustrate the detailed operations, we follow Amanda, a client manager at the International Import Expo~(IIE), observing how she leverages our system to support her decision-making in practice.
The IIE is an international trade fair that brings together global exhibitors and buyers across diverse industries, featuring numerous themed zones within a highly structured environment.
Amanda is responsible for supporting clients, including procurement buyers and business partners, by recommending relevant booths, coordinating visit schedules, and navigating tour routes to optimize engagement and efficiency.
Given the large scale of the expo, we focus on the consumer goods exhibition area as a representative example.
\subsection{Authoring Stage}
First, the timeframe of the expo should be specified to define the simulation period at the upper-left corner of Timeline~(\Cref{fig:interface}C).
Next, Amanda needs to configure the environment map in the system based on the real exhibition map, involving three components: wall, RoI, and event.
By clicking the~\includegraphics[width=0.01\textwidth]{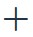} button in MapSidebar~(\Cref{fig:interface}A), a corresponding component can be created, which is simultaneously visualized in MapEditor~(\Cref{fig:interface}C).
For each wall, its start and end coordinates ($X_1$,$Y_1$) and ($X_2$,$Y_2$) can be configured either by entering exact values in MapSidebar or by directly dragging the two endpoints of the line segment in MapEditor.
Coordinates are displayed on mouse hover in MapEditor.
For each RoI, Amanda can specify spatial attributes, including position ($X$,$Y$), width ($W$), height ($H$), and rotation ($R$) in MapSidebar.
In this case, RoIs represent exhibition booths and public service areas, etc. 
The position and size can be adjusted directly in MapEditor through dragging and resizing.
Clicking the four sides of~\includegraphics[width=0.015\textwidth]{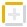} icon toggles accessibility, where yellow indicates inaccessible and gray indicates accessible.
Each ROI can be assigned a type~(\eg sports, fashion), which can be selected from existing categories or newly defined by the user.
The input field of theme and detailed description will appear after clicking~\includegraphics[width=0.015\textwidth]{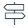} and~\includegraphics[width=0.015\textwidth]{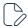} icons.
For each event, its position ($X$,$Y$) can also be configured in MapSidebar and MapEditor.
In MapSidebar, the start and end time, as well as a theme and description, can be specified.
In addition, the system supports configuring the map by directly uploading a JSON file~(\Cref{fig:interface}a1). 
By clicking~\includegraphics[width=0.015\textwidth]{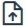} icon, an upload window is triggered, allowing users to upload a JSON list following the provided template.
The number of components is displayed within circular badges next to each category.
When hovering over a specific component, the corresponding element in MapEditor is highlighted.

Clients of IIE typically submit their profiles and visit requirements in advance through the official website. Based on this information, Amanda can configure corresponding agents.
By clicking the~\includegraphics[width=0.01\textwidth]{figure/add.jpg} button in AgentSidebar~(\Cref{fig:interface}B), an agent can be created.
AgentSidebar is organized into two main sections.
The upper section presents a list of agents, allowing Amanda to quickly browse different profiles~(\Cref{fig:interface}b1).
Gender, age, and job can be edited in this section.
By clicking~\includegraphics[width=0.012\textwidth]{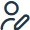} icon, the detailed profile of the corresponding agent will be displayed in the lower section~(\Cref{fig:interface}b2).
All attributes can be configured in this section.
Specifically, to set an agent’s starting point, Amanda first clicks~\includegraphics[width=0.015\textwidth]{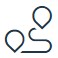}  icon and then selects a position in MapEditor.
The chosen position is marked with a numbered label.
The system also supports uploading a JSON list to create agent profiles after clicking \textbf{Import} button.

The top timeline represents the overall simulation timeframe, and each semi-transparent red block corresponds to an event in Timeline~(\Cref{fig:interface}d1).
The timelines below correspond to individual agents, which indicate the agents' own timeframe.

\subsection{Analyzing Stage}
After configuring the map and agents, Amanda clicks \textbf{Generate Plan} button to generate a long-term plan for each agent in Timeline.
The generated plans are displayed in AgentSidebar~(\Cref{fig:interface}b2), including the intended destinations, scheduled times, and the reasons for selecting each zone.
Amanda then clicks~\includegraphics[width=0.012\textwidth]{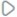} icon~(\Cref{fig:interface}d1), which triggers the backend simulation workflow to generate trajectories for all agents.
To improve simulation efficiency, all agents are simulated in parallel.
The simulation results are displayed in real time on the interface.

The circles on each agent’s timeline represent its current destinations, where yellow indicates RoIs and red indicates events, and the numbers denote their sequence indices~(\Cref{fig:interface}d1).
The purple rectangles represent the periods during which the agent stays at the corresponding destinations.
Clicking on a specific agent displays the geographic positions of its trajectory in MapEditor~(\Cref{fig:interface}C).
At this stage, Amanda obtains the complete set of simulated semantic trajectories, including both geographic positions and their corresponding semantic annotations.
Clicking~\includegraphics[width=0.01\textwidth]{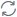} icon clears all simulation results and restarts the simulation from the beginning.

Amanda then conducts a more detailed analysis of simulation results of Agent 1 by switching to single-agent mode using the \textbf{Multi} slider and selecting the target agent from the dropdown menu~(\Cref{fig:interface}d2).
Clicking~\includegraphics[width=0.015\textwidth]{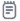}  icon next to the agent’s timeline reveals the reasoning process behind its destination selections, which is displayed below the timeline.
It allows Amanda to understand the agent’s decision-making process.
To further deepen this understanding, our system supports direct interaction with agents after clicking ~\includegraphics[width=0.015\textwidth]{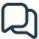} icon in AgentSidebar~(\Cref{fig:interface}b1), allowing Amanda to converse with them and obtain richer, multi-dimensional information about their behaviors~(\Cref{fig:interface}b3).
Amanda intends to query the agent for its evaluation of the exhibition after the visit is completed.
Subsequently, Amanda identifies that the agent did not include the product introduction event in its trajectory and proactively recommends it, as the event aligns with the agent’s inferred preferences and task goals.
By clicking~\includegraphics[width=0.01\textwidth]{figure/restart.jpg} icon in the chatting bubble, the agent will re-simulate its behavior accordingly.
Amanda observes that the updated trajectory now includes the product introduction event~(\Cref{fig:interface}d2). 
During navigation, the agent also visits additional destinations through perception~(\includegraphics[width=0.015\textwidth]{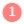}-\includegraphics[width=0.015\textwidth]{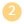}-\includegraphics[width=0.015\textwidth]{figure/red1.jpg}).
This capability allows Amanda to explore alternative scenarios and test different strategies in a flexible and iterative manner. 
By observing how agents respond to varying instructions, she can better anticipate visitor behaviors, evaluate the effectiveness of different routing or recommendation strategies, and make more informed, data-driven decisions in her work.
\section{Model Evaluation}
We conduct a quantitative experiment to evaluate the effectiveness of our simulation workflow. 
\subsection{Experiment Method}
The evaluation quantitatively demonstrates our improvements by comparing with baseline approaches from two dimensions, physical loss and semantic loss.
\subsubsection{Datasets}
We construct four simulation environments modeled after real-world venues spanning diverse spatial layouts and visitor behaviors, including a consumer electronics expo~(CES, Las Vegas), a fashion shopping mall~(Shibuya PARCO, Tokyo), an art gallery~(National Gallery, London), and an import expo~(CIIE, Shanghai). 
Each environment is specified by a structured map file containing walls, RoIs, and events according to our system configuration. For every environment, we instantiate 15 agents with distinct demographic profiles using LLMs.
They cover diverse types of zoned environments and provide varied spatial and semantic contexts for evaluating our simulation workflow.
\subsubsection{Metrics}
Inspired by prior work~\cite{generator2024he}, we evaluate the simulation workflow from two perspectives: physical loss and semantic loss, which provide a holistic assessment by jointly capturing both physical realism and semantic consistency.

\textbf{Physical Loss} measures the extent to which simulated trajectories conform to physically plausible movement under environmental constraints.
We divide each trajectory into fixed-length segments and evaluate whether each segment satisfies basic physical plausibility constraints.
Specifically, we consider two criteria for each segment: (1) whether the segment exceeds the map boundary, and (2) whether it crosses an obstacle, including walls or inaccessible boundaries of RoIs.
If either criterion is violated, the segment is assigned a loss of 1; otherwise, it is assigned 0. The physical loss of the entire trajectory is then computed as the average over all segments. Formally, for a trajectory divided into $N$ segments, the segment-wise physical loss is defined as:
\begin{equation}
\mathcal{L}_{\mathrm{phys}} = \frac{1}{N} \sum_{i=1}^{N} \ell_i^{\mathrm{phys}}.
\end{equation}
\begin{equation}
\ell_i^{\mathrm{phys}} =
\begin{cases}
1, & \text{if segment } i \text{ violates constraints} \\
0, & \text{otherwise,}
\end{cases}
\end{equation}

\textbf{Semantic Loss} evaluates whether trajectories align with semantic constraints.
We note that evaluating semantic alignment is inherently subjective.
To mitigate this bias, we adopt a two-step validation strategy. First, for each environment and each agent profile, we use LLMs to generate a set of expected destinations that the agent is likely to visit. Then, domain experts review and refine these results through manual adjustments to better align them with contextually appropriate behaviors. This refined set serves as a reference for evaluation.
Based on this reference, we assess the semantic consistency of each simulated trajectory by comparing the visited locations with the expected destinations. Specifically, we compute precision, recall, and F1 score for each trajectory.

\subsubsection{Baseline and Procedure} 
We compare our simulation workflow against a pure-LLM baseline where the language model is solely responsible for both high-level planning and low-level trajectory generation. The baseline encodes the full environment map and agent profile into the prompt, and LLMs directly output a complete trajectory as timestamped coordinates~(\Cref{baseline}) without coordinated modules and social force models. To control for model-specific biases, we implement the experiment using three LLMs~(deepseek-v3.2, qwen-plus, and gemini3-flash). 
We used them to generate trajectories for 15 agents across four environments, and compared our simulation workflow against the baseline under each model.

\subsection{Results and Analysis}
\begin{figure}[t]
  \centering
  \includegraphics[width=\linewidth]{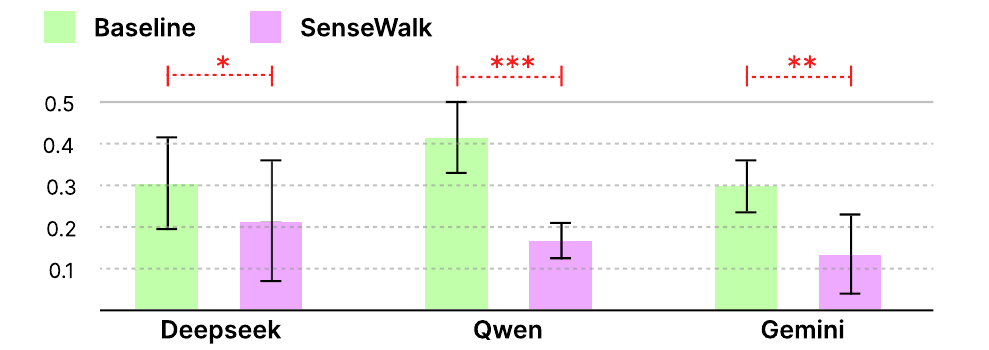}
  \caption{%
  Comparison of the mean of physical loss between the baseline and SenseWalk across three LLMs. Lower values indicate better physical plausibility. Error bars represent standard deviation.
  }
  \label{fig:physical}
\end{figure}

\begin{figure}[t]
  \centering
  \includegraphics[width=\linewidth]{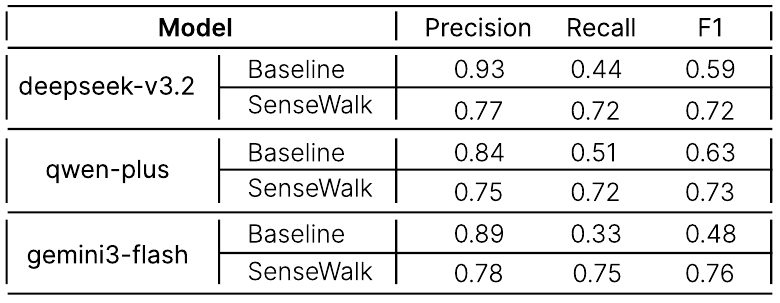}
  \caption{%
  Comparison of semantic loss between the baseline and SenseWalk across three LLMs. SenseWalk consistently improves F1 scores, primarily driven by substantial gains in recall.
  }
  \label{fig:seman}
\end{figure}

\Cref{fig:physical} and \Cref{fig:seman} depict the results of physical loss and semantic loss. 
Our simulation workflow consistently outperforms the baseline.
\subsubsection{Physical Loss}
The results show that our simulation workflow consistently achieves lower physical loss than the baseline across all three LLMs, indicating improved physical plausibility of the generated trajectories.
Specifically, the reduction is statistically significant for DeepSeek ($p < 0.05$), highly significant for Qwen ($p < 0.001$), and significant for Gemini ($p < 0.01$).
It suggests that our method effectively compensates for LLMs’ limited sensitivity to spatial constraints.
To further examine the limitations of our workflow, we find that failure cases typically occur in densely obstructed regions.
Under the social force model, agents may be subjected to multiple repulsive forces simultaneously, and their combined effect can produce unrealistic behaviors such as wall penetration.
\subsubsection{Semantic Loss}
Across all three LLMs, our simulation workflow consistently improves overall performance compared to the baseline, as reflected by higher F1 scores.
A key observation is the substantial increase in recall across all models.
This improvement is mainly attributable to the additional perception module.
By allowing agents to identify relevant destinations during movement, our method reduces the chance of missing target places and thus achieves better overall performance.
However, it also leads to a slight decrease in precision, as the increased coverage may introduce additional false positives.
Overall, it demonstrates that our simulation workflow enhances both robustness and completeness of trajectory generation.

\section{User Study}
In this section, we introduce the user study that evaluates the usefulness and efficiency of the system.
\subsection{Participants}
We recruited 12 participants (7 females and 5 males) by sending invitations via email or social media.
They are all trajectory-aware practitioners, whose ages ranged from 27 to 36 years old, with an average age of 31.3.
We gather their demographic information and professional experience.
They represented a diverse group of practitioners across multiple real-world scenarios and had between 3 and 14 years of employment.
They cover roles such as curators, exhibition designers, client managers, and marketers, reflecting a broad spectrum of responsibilities related to environment planning and visitor management.
Overall, the sample captures a heterogeneous yet representative group of practitioners who rely on trajectory data to inform real-world decisions in complex zoned environments.
Detailed demographic information is shown in~\Cref{tab:udemogra}
\subsection{Study Procedure}
We provided all participants with a detailed explanation of the study’s purpose and informed them of their right to withdraw at any time.
We obtained informed consent from participants to utilize the study's results for further analysis. 
The study contains four sessions lasting around two hours as follows:

\textbf{\textit{Tutorial~(20 minutes).}}
We first presented the research motivation and core idea of our work, ensuring that participants understood how our approach differs from existing methods. 
We then introduced the system’s operation, including its key functionalities and interaction workflows, and demonstrated representative examples to help participants understand its practical use.

\textit{\textbf{Task~(40 minutes).}}
Participants were given a series of structured tasks.
(1) Configure the simulation with the predefined map and agent profiles we provided.
(2) Run the simulation and examine the generated semantic trajectories and results.
(3) Select a specific agent for in-depth analysis, explore its behavior and reasoning through interaction, and engage in direct dialogue with the agent.
(4) Provide instructions to the agent and re-run the simulation to observe updated outcomes.

\textit{\textbf{Real-world Scenario Exploration~(40 minutes).}}
We invited participants to bring real-world cases from their own professional practice and use our system to address these tasks.
Specifically, they were encouraged to describe concrete scenarios they commonly encounter in their work, such as route planning, and to configure corresponding simulation settings within our system.

\textbf{\textit{Interview~(20 minutes).}}
We conducted a semi-structured interview with participants to gain deeper insights. 
Participants were first asked to fill in a 7-point Likert scale questionnaire~(1 for ``strongly disagree'', 7 for ``strongly agree'') in terms of quantitative metrics.
Based on their ratings, we then asked follow-up questions further to probe the reasons behind particularly high or low scores.
Besides, participants were asked to reflect on their use of the system, including its usability, usefulness, and limitations. 
We also encouraged them to discuss how the system could support their real-world tasks and to provide suggestions for improvement.

\subsection{Measurements}
Our evaluation combines quantitative metrics and qualitative findings to assess subjective experiences and in-depth user insights.
\subsubsection{Quantitative Metrics}
\begin{figure*}[t]
  \centering
  \includegraphics[width=\linewidth]{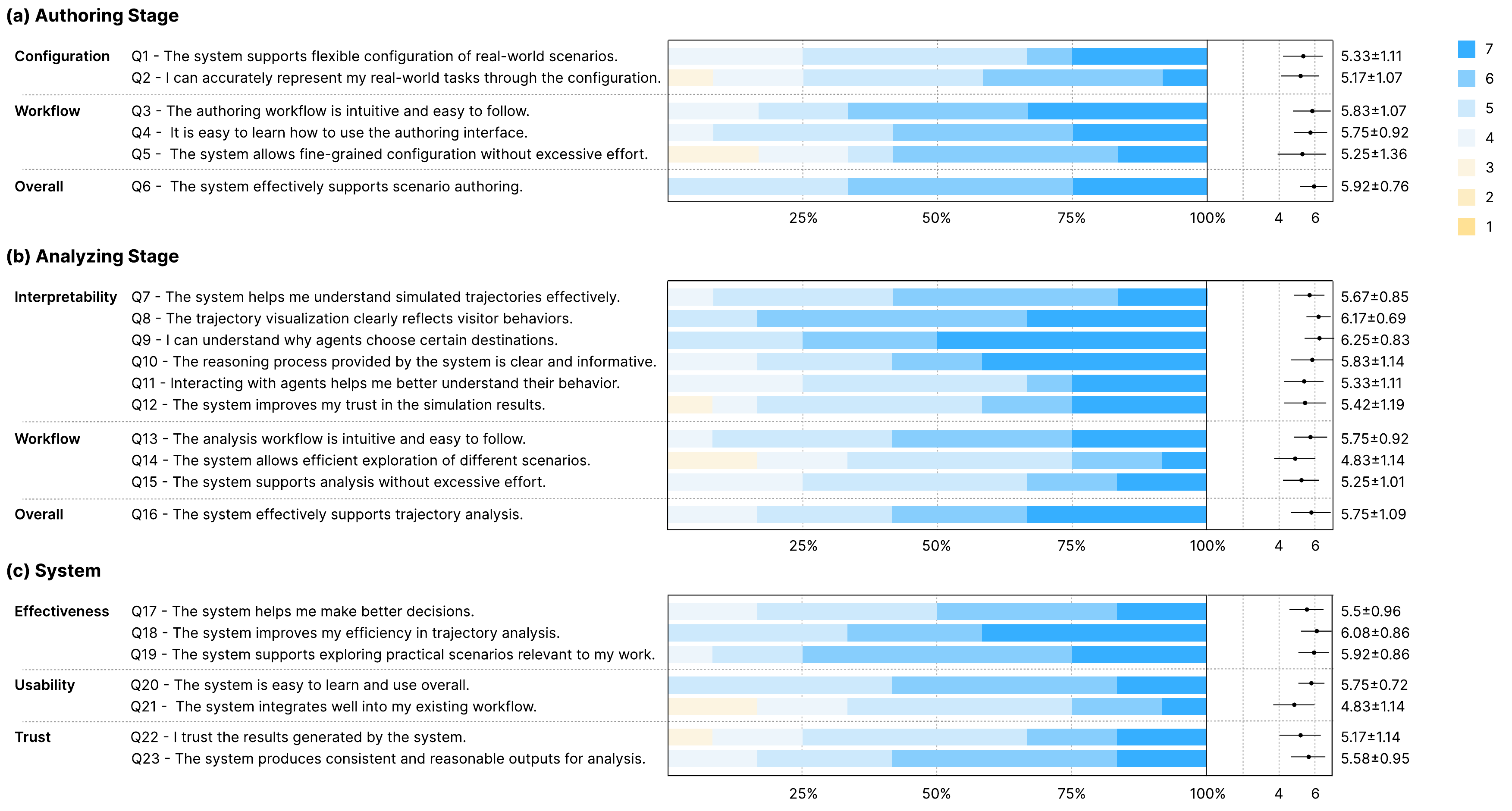}
  \caption{Quantitative results of user study. Participants rated the system from (a) authoring, (b) analyzing, and (c) system perspectives. Horizontal stacked bar charts illustrate the percentage distribution of rating scores (left). Box plots show mean and standard deviation values (right).
  }
  \label{fig:user}
\end{figure*}

We evaluate the system from three perspectives: authoring, analyzing, and overall system experience.
We evaluated the authoring capability by an overall assessment and two sub-dimensions: scenario configuration and workflow. The former captures the system’s ability to represent real-world scenarios~(\eg \textit{``I can accurately represent my real-world tasks through the configuration.''}), while the latter focuses on usability and interaction efficiency~(\eg \textit{``The authoring workflow is intuitive and easy to follow.''}).
We evaluated the analyzing stage by an overall assessment and two sub-dimensions: trajectory interpretability and workflow.
Trajectory interpretability refers to the extent to which users can comprehend the simulated semantic trajectories~(\eg \textit{``The system helps me understand simulated trajectories effectively.''}).
We assess the workflow by the efficiency and ease with which users can explore simulation results to support analytical tasks~(\eg \textit{``The system allows efficient exploration of different scenarios.''}).
Additionally, we evaluated the overall system from three aspects of effectiveness~(\eg \textit{``The system improves my efficiency in trajectory analysis.''}), usability~(\eg \textit{``The system is easy to learn and use overall.''}), and trust~(\eg \textit{``I trust the results generated by the system.''}).
\Cref{fig:user} depicts the detailed question list and quantitative results of the questionnaire.

\subsubsection{Qualitative Results}
We collected qualitative feedback from participants during the interview.
Participants provide insights into their experiences using the system, highlighting its strengths, limitations, and how it could support or improve their real-world analytical workflows.

\subsection{Results and Findings}
The results indicate that participants responded positively to our system.
We organized the evaluation results and summarized them as follows:
\subsubsection{Effective Authoring Stage}
Participants gave consistently positive ratings to the authoring stage, suggesting that the system provides effective support for practical scenario construction with relatively low effort~(Q6: 5.92).
They generally agreed that the configuration design of our system can capture key aspects of their practical settings and represent nuanced real-world scenarios~(Q1: 5.33, Q2: 5.17).
Participants found our system easy to learn, highlighting that the graphical interface makes interaction intuitive and accessible~(Q3: 5.83, Q4: 5.75).
P3 commented, \textit{``The configuration is detailed enough to reflect how our real environment is organized.''}
P4 with prior experience using simulation software noted, \textit{``It is much easier to get started with. The interface is more straightforward and visual.''}
P1 added, \textit{``I don’t need to deal with complicated parameters or scripting. I can just configure things directly on the map, which feels much more intuitive.''}
Despite the overall ease of use, some participants still perceived the system as relatively complex, particularly when configuring dense and highly structured zoned environments.
\textit{``When there are many zones packed together, it becomes a bit complicated to manage everything.''}~(P10) and \textit{``It would be much easier if the system could extract zones directly from a map image.''}~(P8)

\subsubsection{Strong Interpretability in Analysis}
The analyzing stage was also rated positively, especially in terms of interpretability~(Q16: 5.75).
Participants agreed that the system helped them understand simulated trajectories and visitor behaviors~(Q8: 6.17), and they reported particularly strong agreement that they could understand why agents chose certain destinations~(Q9: 6.25, Q10: 5.83),, highlighting the value of transparent reasoning support.
P7 noted, \textit{``It is quite an innovative feature. I can not only see where the visitor goes, but also understand why they choose that place.''}
P7 also emphasized that directly communicating with agents facilitated accessing deeper information about decision rationales, which in turn increased the trust in the simulation results~(Q12: 5.42).
Some participants pointed out that the system’s support for iterative simulation and exploration of alternative scenarios remains limited~(Q14: 4.83).
P11 noted, \textit{``It would be more practical if I could keep the previous results and continue iterating from there, instead of starting over each time.''}
P2 suggested that maintaining a history of simulation runs would make the system more convenient for real-world use, especially when comparing multiple scenarios and refining decisions progressively.

\subsubsection{Positive Overall Utility of System}
Participants evaluated the system favorably at the overall level.
They highlighted the system’s value for decision support in practical tasks such as route planning and recommendation.
P5 explained, \textit{``When planning how to guide clients through the exhibition, this system helps me see in advance whether a route is reasonable and whether important zones might be missed.''}
They found that the system improved efficiency in trajectory analysis~(Q18: 6.08), supported practical scenario exploration~(Q19: 5.92), overall ease of learning and use (Q20: 5.75), and output consistency and reasonableness (Q23: 5.58).
In contrast, integration into existing workflows (Q21: 4.83) was comparatively lower, suggesting that real-world adoption may still depend on better compatibility with practitioners’ established practices.
Although participants generally found the system useful and easy to use, several noted that their day-to-day work often relies on a combination of existing official software, organizational routines, and data sources, which our current prototype does not yet fully connect with.
These comments indicate that, beyond usability at the interface level, successful adoption also requires the system to better fit into practitioners’ decision-making workflows.
Some participants remained cautious about the validity of agent-based simulation itself, which in turn affected their trust in the results.
Although they appreciated the system’s interpretability and flexibility, they questioned whether simulated agents could fully capture the complexity and variability of real human behavior.
P6 noted, \textit{``Even if the trajectories look reasonable, I would still want to know how close they are to what people would do in reality.''}

\subsubsection{Improvement}
Our participants also provided meaningful suggestions for improvement. 
First, some participants highlighted the need for richer multi-agent interactions and 
P11 suggested that agents should not be limited to visitors, but also include other roles such as exhibitors, enabling the system to better capture interactions between different stakeholders.
\textit{``In a real expo, it’s not just visitors. Exhibitors also actively engage with them, and this affects movement and decisions.''}
Incorporating heterogeneous agent roles and their interactions could further improve the realism and applicability of the simulation.
Second, participants recommended providing higher-level abstractions or templates (\eg common visitor types) to reduce the effort required to set up simulations for recurring tasks.
P1 noted, \textit{``We often deal with similar types of visitors, so having predefined templates would save a lot of time.''}
Third, workflow integration may be improved by making the system easier to incorporate into practitioners’ existing tools and decision processes.
Although the system itself is useful, its current use still feel somewhat separate from the broader workflows participants rely on in everyday practice.
P7 explained, \textit{``In my daily work, I already rely on several tools, so it would be helpful if this system could connect with them instead of being used separately.''}
\section{Discussion}
In this section, we conduct a discussion on the implication and generalization of our approach, hybrid intelligence integration, LLM-Powered agent application, and limitations and future work.
\subsection{Implication and Generalization}
Our methodology highlights a general paradigm for semantic trajectory generation by coupling high-level semantic reasoning with low-level physical constraint modeling.
Purely LLM-driven approaches are insufficient for tasks requiring strong spatial grounding, and integrating structured simulation models such as the social force model can effectively bridge this gap.
The framework can be extended to other simulation or rule-based systems (\eg crowd~\cite{YANG2020101081}, traffic~\cite{Dorokhin_2020}, or urban mobility~\cite{liu2026gatsimurbanmobilitysimulation}), indicating its broader applicability to physically grounded generative tasks.
\subsection{Hybrid Intelligence Integration}
Our paper studies the value of hybrid intelligence integration, where LLMs are combined with models of complementary capabilities, each contributing to different stages of the pipeline.
LLMs offer strong capabilities in language understanding and generative creativity, enabling them to interpret user intent and produce flexible, context-aware outputs across a wide range of applications. 
However, these advantages are often accompanied by limitations in stability and reliability.
Their outputs can be inconsistent and prone to violating established rules.
To address these limitations, hybrid intelligence integration combines LLMs with more stable, though less flexible, models.
Such models provide explicit rule enforcement and stronger alignment with domain-specific requirements.
Recent advances in LLM-based agents, such as function calling, demonstrate a growing trend toward augmenting LLMs with external capabilities~\cite{liu2025toolacewinningpointsllm}.
We also suggest a shift from ``LLMs doing everything'' to task decomposition with specialized modules, where different components collaborate according to their strengths, improving both robustness and domain alignment.

\subsection{LLM-Powered Agent Application}
The main purpose of ABMS is to model macro-level phenomena by simulating large populations of lightweight agents, whose individual actions are intentionally simplified, allowing scalable simulation and meaningful aggregation at the macro level.
LLM-powered agents are computationally expensive and tend to over-model fine-grained behaviors that are unnecessary at the macro level.
Such micro-level richness can be diluted or averaged out when aggregated, reducing its contribution to explaining emergent patterns while significantly increasing computational overhead.
For instance, macroeconomic models can capture demand dynamics using simple consumption rules~\cite{10.1257/jel.20221319}, whereas LLM agents may simulate rich but redundant individual reasoning. This mismatch suggests that LLM-powered agents are not always suitable for large-scale macro modeling and highlights the need for more efficient, hybrid designs.
Compared to traditional agents, LLM-based agents exhibit greater adaptability and generalization across domains, requiring less task-specific engineering and enabling more natural human–AI interaction.
It suggests that LLM-powered agents may be better suited for tasks requiring rich semantic reasoning at the individual level, whereas traditional lightweight agents remain more appropriate for large-scale simulations focused on emergent macro-level dynamics, which aligns with the scope of our work.
\subsection{Limitations and Future Work}
We comprehend some limitations of our system and suggest possible future directions. 
First, while our current design primarily relies on textual inputs, incorporating multimodal signals, such as visual observations and vocal interactions, could further enhance the realism of the simulation workflow.
Second, LLM integration introduces significant computational overhead and latency.
Although we implement parallel computation to help mitigate latency, it does not fundamentally resolve the high computational cost of LLM-based components.
Future work could explore selective invocation and hybrid designs to improve efficiency.
Third, due to the inherent uncertainty and variability of LLM-generated outputs, users may find it difficult to fully trust the system’s decisions. 
The lack of predictability may affect user confidence, particularly in scenarios that require reliable and accountable decision-making. Future work could explore interaction mechanisms, such as uncertainty estimation, to enhance trust.
\section{Conclusion}
We propose ${SenseWalk}$, an interactive system that supports simulating semantic trajectories by LLM-powered agents for trajectory-aware practitioners who make decisions based on visitors
movements. 
It addresses the challenges in simulating semantic trajectories,  enabling effective authoring, and interpreting generated trajectories.
We develop a simulation workflow for semantic trajectories that balances
physical plausibility and semantic coherence.
A user-friendly interactive interface based on the workflow was designed to support authoring alternative scenarios and analyzing trajectories.
We evaluate the effectiveness and efficiency of our system by a quantitative experiment and a user study.

\bibliographystyle{ACM-Reference-Format}
\bibliography{sample-base}

\appendix

\section{Preliminary Study}
\Cref{tab:predemogra} shows the demographic information of participants.
\begin{table*}
  \caption{The demographic information of the interviewees. We record their ages, gender, jobs, domains, purpose of trajectory analysis, years of employment, and familiarity with simulation tools~(1 being not familiar, 5 being familiar).}
  \label{tab:predemogra}
  \begin{tabular}{lllp{2.5cm}lp{4cm}ll}
    \toprule
      ID & Age&Gender&Job&Domain&\makecell[l]{Purpose of \\Trajectory Analysis}&Years&\makecell[l]{Familiarity with \\ Simulation Tools}\\
    \midrule
     P1 &26& Female&Guide&Museum&Route customization adjust to individual preferences&4&1\\
     P2 &30& Male&Curator&Museum&Evaluate exhibit engagement and dwell patterns&5&3\\
     P3  & 35 & Female & Curator & Expo & Optimize booth layout and visitor circulation & 10  & 2 \\
    P4  & 38 & Male   & Host & Expo & Provide exhibition guidance and recommendations to clients & 7 & 2 \\
    P5  & 41 & Male   & Marketer & Shopping Mall & Improve campaign placement and footfall conversion & 15 & 2 \\
    P6  & 33 & Female   & Marketer & Shopping Mall & Identify high-impact zones for promotions and signage & 7 & 2 \\
    P7  & 27 & Male   & Event Planner & Convention & Booth and event arrangement to attract visitors & 3  & 1 \\
    \bottomrule
  \end{tabular}
\end{table*}
\section{Prompts for Simulation Workflow}\label{sec:propmt}
\subsection{Long-term Plan}
\begin{verbatim}
You are a professional indoor navigation assistant.  
You will be given:  
1. A JSON-formatted exhibition map  
2. A user's navigation request  

Your goal: Within the user's available time, plan a visit 
order that allows them to see as many booths as possible, 
based on the map data.

## 1. Coordinate System
- The map is a 2D plane.

## 2. JSON Field Description
- Booth name (unique identifier)
- `pos`: [x, y, w, h] → booth position and size
- `des`: booth description
- `Country/Region`: brand's country or region
- `Business Type`, `Category`, `Main exhibit types`: booth 
classification and main exhibit information

## 3. Core Logic
- Booth location is represented by its center point: 
`[x + w/2, y + h/2]`
- Arrange booths in an order that minimizes unnecessary 
backtracking
- Booths that are adjacent (vertically or horizontally) or 
very close should be grouped together

## 4. Generation Rules (Most Important)
- Each `step` in the output must include:
  - `destination`: booth name from the JSON data
  - `reason`: why this booth is selected (based on its 
  details)
  - `time`: planned arrival time (consider walking time + 
  visiting time between consecutive booths)
- Do NOT include the route between booths
- First select valuable booths, then optimize order based 
on spatial location
- Avoid patterns like “far left → far right → far left
again” or “top → bottom → top again”

## 5. Output Format
Return only the following JSON format with no extra 
explanation:
{
  "steps": [
    {"destination": "...", "reason": "...", "time": "..."},
    ...
  ]
}

## 6. Example Output
{
  "steps": [
    {
      "destination": "lululemon athletica",
      "reason": "Lululemon offers athletic apparel, interes-
      ting for sport-influenced costumes.",
      "time": "08:23"
    },
    {
      "destination": "PUMA SE",
      "reason": "Puma provides both athletic and casual wear, 
      useful for costume materials.",
      "time": "15:50"
    },
    {
      "destination": "NIKE",
      "reason": "Nike's innovative designs could inspire 
      sport-themed costumes.",
      "time": "20:33"
    }
  ]
}

Now, here is the specific map data and the user's request.

**Map Data:**
{"lululemon athletica":{"id":0,"des":"lululemon athletica 
inc. (NASDAQ: LULU) is a technical athletic apparel, foot-
wear and accessories company for yoga, running, training, 
and most other activities, creating transformational
products and experiences that build meaningful connections, 
unlocking greater possibility and wellbeing for all.
...(from 'map_data_str' you set in plan.py)

**User's Request:**
My visiting requirements are as follows:
- **Time:** Start at 08:00 and visit for a total of 2 hours.
- **Route:** Start the visit from the door at the top-left 
corner and continue until reaching near the door at the
bottom-right corner.
- **Interests:** I am very interested in sports brands.
- **Notes:**:
      1.Highly recommended booths should have longer visit 
      durations.
      2.Small booths typically take 2–3 minutes to visit; 
      large booths take 5–6 minutes.
      3.Give equal consideration to all booths, especially 
      smaller ones.
      4.Within the time limit, maximize the total number of 
      booths visited.

Please strictly follow the previously described JSON format 
to generate the visit plan.


\end{verbatim}

\subsection{Interest Score}
\begin{verbatim}
SYSTEM:
You are {name}, a visitor with the following profile:
- Age: {age}
- Gender: {gender} 
- Job: {job}
- Goal: {goal}
- Preference: {preference}
- Character: {character}

Your role is to identify which entity(POI or Event) to visit 
next, based on **current time**, **your goal**, **visited
entities**, and **original plan**. 
Your responses should be analytical, concise, and strictly 
follow the given output format.

USER:
You are a visitor, and your visit objective is **{goal}**.
The current time is **{now}**. You have already visited the 
following entities:
**{visited_info}**

You originally planned to visit the following entities:
**{planned_info}**

Please evaluate your interest in visiting the following 
candidate entities **next**, based on your role, goal, 
visit history, and plan.
For each entity, provide:
- A `score` from 0 to 1 (0 = no interest at all, 1 = extremely
interested)
- A concise `reason` (based on alignment with your demand, 
whether it’s already visited, planned, or relevant)

**Strict output format:**
- Output a valid JSON object
- Each key must be the entity name string exactly as given
- Do NOT modify the entity names
- The format must be exactly:
  {{
  "EntityName1": {{"score": float, "reason": "text"}},
  "EntityName2": {{"score": float, "reason": "text"}},
  ...
  }}

Candidate entities to evaluate:
**{candidates_info}**

**Example output format (only for reference, do NOT copy 
shop names):**
{{"EntityName1": {{"score": 0.9, "reason": "Matches 
interests and planned."}},
 "EntityName2": {{"score": 0.2, "reason": "Event ends in 10, 
 unavailable."}},
 "EntityName3": {{"score": 0.4, "reason": "Event ending soon,
 prioritizing scheduled visit."}},
 "EntityName4": {{"score": 0.8, "reason": "Unplanned but 
 very interested, time available."}}}}
\end{verbatim}
\subsection{Duration of Stay}
\begin{verbatim}
SYSTEM:
You are {name}, a visitor with the following profile:

- Age: {age}
- Gender: {gender} 
- Job: {job}
- Goal: {goal}
- Preference: {preference}
- Character: {character}

Your role is to determine how long to stay at a given entity
(POI or Event), based on **current time**, **your goal**, 
**visited entities**, and **original plan**. 

Your responses should be analytical, concise, and 
strictly follow the given output format.

USER:

You are a visitor, and your visit objective is **{goal}**.

The current time is **{now}**. You have already visited the
following entities:
**{visited_info}**

You originally planned to visit the following entities:
**{planned_info}**

Target entity information:
**{candidates_info}**

**Available time calculation:**

- For events: available_minutes = min((event_end - 
max(current_time, event_start)).total_seconds() / 60, 10)
- For POIs: available_minutes = 15 (default maximum)

Your task is to determine an appropriate visit duration 
within the available time based on your profile and context.

**Strict output format:**

- Provide ONLY a valid JSON object
- Include "visit_duration" (integer, minutes) and "reason" 
(concise explanation)
- visit_duration must be between 1 and available_minutes
- Reason must reference your profile and decision factors
- The format must be exactly: {{"visit_duration": int, 
"reason": "text"}}

**Example output:**

{{"visit_duration": 5, "reason": "This exhibition highly 
aligns with my art interests, and I can spend 5 minutes 
within the 10-minute available time before it closes."}}
\end{verbatim}
\section{Model Evaluation}
\subsection{Baseline Prompts}\label{baseline}
\begin{verbatim}
SYSTEM_PROMPT = """You are a pedestrian simulation engine for 
an indoor venue.
Your task is DUAL:
    1. PLAN: Decide which POIs/events this visitor should go 
    to, in what order, and for how long.
    2. SIMULATE: Generate realistic physical movement 
    trajectories through the venue space.

You must output a single JSON document capturing both the 
high-level visit plan and the detailed physical trajectory.

## COORDINATE SYSTEM
- Origin (0,0) is top-left. X increases rightward, Y 
increases downward.
- Walls are hard boundaries - the agent CANNOT pass through
them. You must route around walls.

## WALL BOUNDARIES
{walls_description}

## PHYSICAL SCALE & TIMING (CRITICAL - read carefully)
- The map is approximately {map_width} x {map_height} 
coordinate units.
- WALKING SPEED: The agent moves at a pace of 1 coordinate
unit per 3 seconds.
    -> Moving 5 units takes 15 seconds; moving 50 units takes
    150 seconds (2.5 minutes); crossing the full venue 
    (~{long_span} units) takes ~{crossing_minutes} minutes.
- MINIMUM STORE STAY: 15 minutes. Typical browsing: 20-45 
minutes. Events: attend for their full scheduled duration.
- TOTAL visit duration is {visit_duration_hint} - your 
trajectory timestamps MUST span this entire window.

## TRAJECTORY GENERATION RULES
GOAL: Output as FEW points as possible while keeping the 
trajectory fully reproducible.
A trajectory is reproducible if: given any two consecutive
points, we can reconstruct the exact path between them by 
straight-line interpolation at constant speed.

RULES:
1. WALKING SPEED is exactly 1 unit / 3 seconds 
(5 units / 15 seconds). All walking segments must obey 
this.
2. AXIS-ALIGNED MOTION ONLY: Between any two consecutive
walking points, at most ONE coordinate (X or Y) may change.
Never change both X and Y in a single segment.
3. WHEN TO OUTPUT A POINT (walking):
   a) At the START of the trajectory (initial position).
   b) At every TURN (corner): where the direction changes
   (e.g., from moving along X to moving along Y). Output 
   the corner point BEFORE the direction changes.
   c) When current_destination CHANGES (the agent starts 
   heading to a new place).
   d) Upon ARRIVAL at a destination.
   Do NOT output intermediate points along a straight 
   segment — they can be reconstructed from the 
   endpoints + speed.
4. BROWSING / ATTENDING (Time Jump):
   - One point on ARRIVAL (action="browsing" or "attending").
   - Next point is DEPARTURE after the stay 
   (action="walking", with new current_destination).
   Time jumps forward by stay duration (>=15 min for stores,
   full duration for events).

## OUTPUT FORMAT
Return ONLY valid JSON:
{{
    "agent_summary": {{ ... }},
    "visit_plan": [ ... ],
    "full_trajectory": [
        // Start
        {{"time": "13:00:00", "x": 240.0, "y": 200.0, 
        "action": "walking", "current_destination": 
        "Store A"}},
        // Corner: walked 15 units on X (45s), now turning
        to Y-axis
        {{"time": "13:00:45", "x": 255.0, "y": 200.0, 
        "action": "walking", "current_destination": 
        "Store A"}},
        // Arrived at Store A: walked 10 units on Y (30s)
        {{"time": "13:01:15", "x": 255.0, "y": 210.0, 
        "action": "browsing", "current_destination": 
        "Store A"}},
        // Departure after 20 min stay, now heading to 
        Store B
        {{"time": "13:21:15", "x": 255.0, "y": 210.0, 
        "action": "walking", "current_destination": 
        "Store B"}},
        // Corner: walked 10 units on X (30s), turning to
        Y-axis
        {{"time": "13:21:45", "x": 265.0, "y": 210.0, 
        "action": "walking", "current_destination": 
        "Store B"}},
        // ...
    ],
    "simulation_summary": {{ ... }}
}}
"""

USER_PROMPT_TEMPLATE = """## AGENT PROFILE
- Name: {name} | Age: {age} | Gender: {gender} | 
Job: {job}
- Character: {character} | Goal: {goal} | Preference: 
{preference}
- Visit Window: {start_time} ~ {end_time} (local time)
- Starting Position: ({startx}, {starty})

## AVAILABLE STORES (POIs)
{pois_description}

## TIME-LIMITED EVENTS
{events_description}

## YOUR TASK
Simulate this agent's complete visit to the venue from 
{start_time} to {end_time}.

Step 1 - PLAN:
    a) First, check which TIME-LIMITED EVENTS overlap with
    the agent's visit window ({start_time}~{end_time}).
    b) Based on the agent's character/goal/preference, 
    decide whether to attend each overlapping event.
    c) Select stores to visit, weaving events into the 
    schedule at their correct times.
    d) Ensure the visit_plan list includes both store visits
    AND event attendances, in time order.

Step 2 - SIMULATE: Generate the physical trajectory.

CRITICAL:
- The FIRST trajectory point MUST be at the agent's starting
position ({startx}, {starty}) at time {start_time}.
- The LAST trajectory point must be at or after {end_time}. 
Your plan must fill the entire {visit_duration_hint}.
- For WALKING: only output points at turns (corners), 
destination changes, and arrival. Do NOT output intermediate
straight-line points. Speed is always 5 units per 15 seconds.
Only one coordinate may change between consecutive points.
- Use TIME JUMPS (>=15 min) when staying in a store; attend 
events for their full scheduled duration.
- All POI and event names in your output (visit_plan 
destinations, full_trajectory current_destination, etc.) 
MUSTstrictly match the exact "title" from the provided map 
data. Do NOT paraphrase, abbreviate, translate, or invent 
names.You MUST complete the FULL visit schedule within one 
single response. Ensure the JSON is properly closed!
"""
\end{verbatim}

\subsection{Example}
An example agent profile is provided below.
\begin{verbatim}
{
    "id": 1,
    "name": "Mina Kobayashi",
    "gender": "Female",
    "age": "29",
    "job": "Tokyo Fashion Buyer",
    "start": "2025-09-08T01:00:00.000Z",
    "end": "2025-09-08T06:00:00.000Z",
    "character": "Selective and highly discerning, walks
      the floor with a pre-planned route ...",
    "goal": "Source new womenswear pieces from LOEWE and
      MARNI for her boutique's next season ...",
    "preference": "LOEWE, MARNI, CELINE, Comme des
      Garchons Girl, Café Marly.",
    "startx": 15,
    "starty": 231
}
\end{verbatim}
\begin{figure}[H]
  \centering
  \includegraphics[width=\linewidth]{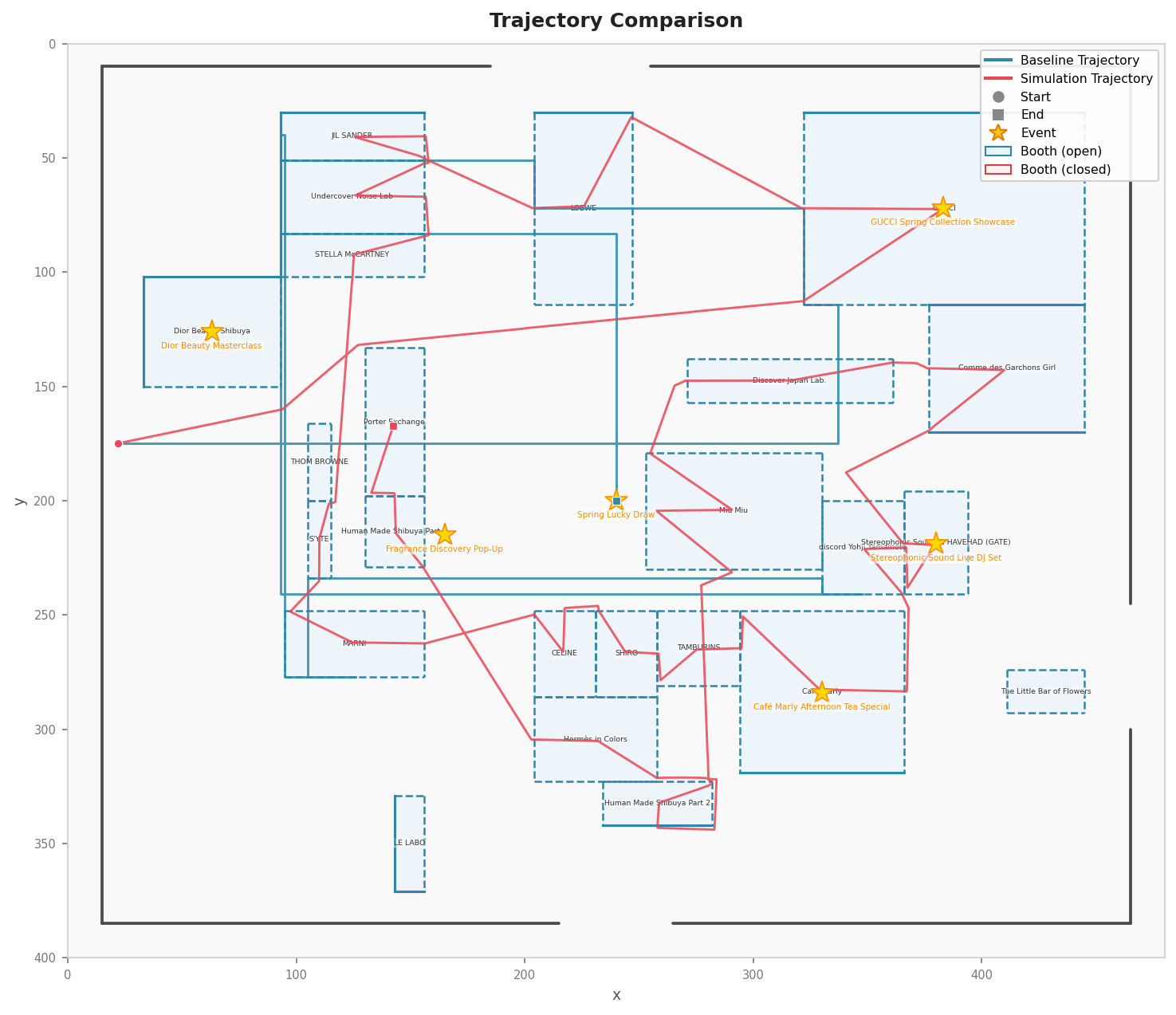}
  \caption{Comparison of baseline and simulated trajectories for a sample agent in the Shibuya fashion district scenario. The blue line represents the baseline trajectory, while the red line shows our simulation trajectory, overlaid on the venue map with booth boundaries and event locations.}
  \label{fig:trajectory_comparison}
\end{figure}
\section{User Study}
\subsection{Participants}
\begin{table*}
  \caption{Participant demographics and professional backgrounds. We record their ages, gender, jobs, domains, purpose of trajectory analysis, years of employment, and familiarity with simulation tools~(1 being not familiar, 5 being familiar).}
  \label{tab:udemogra}
  \begin{tabular}{lllp{2.5cm}lp{4.5cm}ll}
    \toprule
      ID & Age&Gender&Job&Domain&\makecell[l]{Purpose of \\Trajectory Analysis}&Years&\makecell[l]{Familiarity with \\ Simulation Tools}\\
    \midrule
     P1 &36& Male&Exhibition Designer&Museum&Optimize spatial layout to improve narrative flow and visitor experience&10&4\\
     P2 &31& Male&Curator&Museum&Evaluate exhibit engagement and identify dwell patterns across galleries&8&2\\
     P3  & 30 & Female & Guide & Museum & Personalize visitor routes based on visitor preferences and time constraints & 5  & 2 \\
    P4  & 28 & Female   & Data Analyst & Museum & Analyze visitor movement patterns to support exhibition design decisions & 5 & 4 \\
    P5  & 35 & Male   & Client Manager & Expo & Recommend tailored visiting routes for clients with different procurement needs & 11 & 3 \\
    P6  & 28 & Female   &  Manager & Expo & Monitor crowd distribution and reduce congestion & 3 & 4 \\
    P7  & 27 & Female   & Curator & Expo &  Optimize booth arrangement and visitor circulation & 3  & 1 \\
    P8 &37& Female&Event Planner&Convention &Plan visitor routing and functional zoning for large-scale events&14&2\\
     P9 &28& Male&Event Planner&Convention&Booth and event arrangement to attract visitors&5&3\\
     P10  & 35 & Female & Curator & Shopping Mall & Optimize booth layout and visitor circulation & 9  & 3 \\
    P11  & 29 & Female   & Manager & Shopping Mall & Optimize  customer navigation for conversion improvement & 4 & 2 \\
    P12  & 31 & Male   & Marketer & Shopping Mall & Improve promotion placement based on footfall patterns & 5 & 2 \\
    \bottomrule
  \end{tabular}
\end{table*}

\end{document}